\documentclass[12pt,preprint]{aastex}

\begin{document}

\title{GRB 060218/SN 2006aj: prompt emission from inverse-Compton scattering
of shock breakout thermal photons}
\author{Z. G. Dai$^{1,2}$, Bing Zhang$^2$, and E. W. Liang$^{2,3}$}
\affil{$^1$Department of Astronomy, Nanjing University, Nanjing 210093, China\\
$^2$Department of Physics, University of Nevada, Las Vegas, NV 89154,
USA\\ $^3$Department of Physics, Guangxi University, Nanning 530004,
China }

\begin{abstract}
The $\gamma$-ray burst (GRB) 060218/SN 2006aj is a peculiar event, with the
second lowest redshift, low luminosity, long duration, chromatic lightcurve
features, and in particular, the presence of a thermal component in the X-ray
and UV-optical spectra. Thanks to detailed temporal and spectral coverage of
the {\em Swift} observatory, the abundant data allow the GRB prompt emission to
be modelled in great detail for the first time. The low flux of prompt
UV/optical emission disfavors the conventional internal shock/synchrotron
radiation models, which generally predict strong UV/optical emission. Here we
show that the unusual prompt emission of GRB 060218 can be produced by
inverse-Compton scattering of shock-accelerated relativistic electrons off the
detected thermal photons. A pair of (forward plus reverse) shocks form when a
relativistic outflow interacts with a preexisting slower shell. The observed
$\gamma$-ray emission and X-ray emission arise from the reverse-shocked and
forward-shocked regions, respectively. A fit to the data requires an initially
increasing outflow luminosity, which is consistent with the prediction of the
popular collapsar model.
\end{abstract}


\section {Introduction}

The $\gamma$-ray burst (GRB) 060218 was detected by the Burst Alert Telescope
(BAT) onboard {\em Swft} satellite on 2006 February 18.149 UT (Cusumano et al.
2006). The prompt $\gamma$-ray emission lasted about 2000 s and thus it is one
of the longest GRBs. Spectroscopic measurements after the burst revealed that
this event was associated with a type-Ic supernova, SN 2006aj (Modjaz et al.
2006; Pian et al. 2006; Sollerman et al. 2006; Mirabal et al. 2006; Cobb et al.
2006) and that its redshift is $z=0.0331$ (Mirabal \& Halpern 2006). This is
the second lowest redshift GRB after GRB 980425/SN 1998bw. The burst has
several unusual $\gamma$-ray and X-ray properties (Campana et al. 2006, see
Fig.1). First, both prompt $\gamma$-ray emission and X-ray emission detected by
BAT and the X-ray Telescope (XRT) exhibit a slow brightening and a final steep
decay, unlike spiky light curves of classical bursts at higher redshifts. The
lightcurves are chromatic, with the gamma-rays fading up earlier than X-rays.
Second, the peak $\gamma$-ray flux is as low as $\sim 10^{-8}\,{{\rm erg}\,{\rm
cm}^{-2}\,{\rm s}^{-1}}$ and the isotropic-equivalent $\gamma$-ray energy is a
few times $10^{49}$ ergs, about three orders of magnitude lower than the
typical energy of classical GRBs. Third, a soft thermal component was detected
by XRT $\sim 300$ to $10^4$ s after the burst, which a luminosity $\sim
10^{46}\,{{\rm erg}\,{\rm s}^{-1}}$. The thermal component is also identified
in the UVOT data at $\sim 10^5$ s. This soft component is likely due to the
breakout of a radiation-dominated shock, possibly driven by a cocoon
(M\'esz\'aros \& Rees 2001; Ramirez-Ruiz et al.  2002; Zhang et al. 2003),
parts of uncorked envelope matter (Waxman \& M\'esz\'aros 2003) or parts of
accelerated envelope matter (Colgate 1974; Tan et al. 2001). Fourth, the prompt
optical/UV emission is extremely faint and its flux is about five orders of
magnitude fainter than that of the prompt X-ray emission. Finally, it is
interesting to note that the observed X-ray afterglow decayed as a simple power
law with time ($\sim t^{-1.2}$) up to 10 days after the burst.

In the conventional internal shock models (for reviews see Zhang \&
M\'esz\'aros 2004 and Piran 2005), collisions among relativistic shells with
different Lorentz factors would lead to internal shocks and the prompt emission
of a GRB could arise from shock-accelerated electrons by synchrotron radiation
and/or inverse-Compton scattering (e.g. Rees \& M\'esz\'aros 1994; Dai \& Lu
2002; Zhang \& M\'esz\'aros 2002). In these models, the typical synchrotron
self-absorption frequency is in the optical/UV band, so the prompt optical/UV
emission flux is about three orders of magnitude fainter than the prompt X-ray
flux because the synchrotron spectrum is $\nu F_\nu\propto \nu^{4/3}$ in the
optical/UV to X-ray bands. The models predict a strong prompt optical/UV flash
(M\'esz\'aros \& Rees 1999), especially in low-redshift bursts. The tight
constraints of the prompt UV-optical data of GRB 060218 therefore implies that
the conventional internal shocks/synchrotron radiation models are inapplicable
at least for this burst.

In this Letter, we propose a model in which a relativistic outflow interacts
with a preexisting slower shell, leading to a pair of relativistic (forward
plus reverse) shocks. The slower shell could have been a part of the stellar
envelope ejected from the central engine prior to the GRB or a circum-burst
dense clumpy cloud. We show that the unusual prompt emission of GRB 060218
could be produced by inverse-Compton (IC) scattering of shock-accelerated
electrons off the detected shock breakout thermal photons. To our knowledge,
this is the first detailed theoretical model of the prompt emission of a GRB,
which is made possible thanks to the broad temporal and spectral coverage of
GRB 060218 by the {\em Swift} satellite. In \S 2 we present the IC spectrum of
a relativistic shock in a general case. In \S 3 we discuss the shock dynamics
and calculate the emission luminosity and in \S 4 we constrain the model
parameters for GRB 060218. Our results are summarized in \S 5.

\section{IC spectrum in relativistic shocks}

We first discuss a general case in which a relativistic shock with bulk Lorentz
factor of $\gamma$ occurs simultaneously with a soft photon source. We assume
that the luminosity of this photon source is $L_0$ and its radius is much
smaller than the shock radius $R$. We now consider IC scattering of the
shock-accelerated electrons with the soft photons. If the shock expands
outwards isotropically, there should be many head-on scatterings between the
electrons and the soft photons so that some electrons would quickly cool down,
as argued by Brunetti (2000), Wang, Li \& M\'esz\'aros (2006), Wang \&
M\'esz\'aros (2006), and Fan \& Piran (2006). Letting $\epsilon_0$ be the
typical energy of the soft photons in the observer's frame, we have the IC
power and characteristic energy from a relativistic electron with Lorentz
factor $\gamma_e$,
\begin{equation}
P(\gamma_e)= \sigma_Tc\gamma^2\gamma_e^2U'_0,
\end{equation}
\begin{equation}
\epsilon_\gamma(\gamma_e)= \gamma_e^2\epsilon_0.
\end{equation}
The factor $\gamma^2$ in eq. (1) is introduced to transform the power in the
shock rest frame to the observer's frame, $\sigma_T$ is the Thomson cross
section, and $U'_0=L_0/(4\pi R^2\gamma^2c)$ is the energy density of the soft
photons in the shock rest frame. If eq. (1) is multiplied by a factor of $4/3$,
one obtains the IC power of a relativistic electron in an isotropic photon
field. In the present case, a beam of soft photons illuminates the
shock-accelerated electrons along the radial direction. One needs to introduce
a correction factor to the IC power in an isotropic photon field (Fan \& Piran
2006). This correction combined with the factor $4/3$ gives a coefficient of
order unity, which is neglected in eq. (1). The peak spectral density (i.e.,
power per unit energy) is approximated by
\begin{equation}
P_{\epsilon_\gamma,{\rm max}}=
\frac{P(\gamma_e)}{\epsilon_\gamma(\gamma_e)}=
\frac{\sigma_T}{4\pi R^2}\frac{L_0}{\epsilon_0},
\end{equation}
which is independent on $\gamma_e$ and (apparently) on $\gamma$. This
expression represents the number of scatterings with soft photons per unit time
for a certain electron in the Thomson limit. Thus, the peak spectral luminosity
is given by
\begin{equation}
L_{\epsilon_\gamma,{\rm max}}=N_eP_{\epsilon_\gamma,{\rm max}} =
\frac{N_e\sigma_T}{4\pi R^2}\frac{L_0}{\epsilon_0},
\end{equation}
where $N_e$ is the total number of electrons.

Similar to Sari, Piran \& Narayan (1998), if a relativistic electron loses its
energy through the IC discussed here, the critical Lorentz factor $\gamma_c$
above which electrons cool in the observer's time $t$ is defined through
$\gamma\gamma_cm_ec^2=P(\gamma_c)t$, i.e.
\begin{equation}
\gamma_c= \frac{m_ec}{\sigma_T\gamma U'_0t}=\frac{4\pi m_ec^2\gamma
R^2}{\sigma_TL_0t}.
\end{equation}
As usual, we assume that the instantaneous shock-accelerated electron energy
distribution follows a power law: $dN_e/d\gamma_e\propto \gamma_e^{-p}$ for
$\gamma_e\ge \gamma_m$. If a constant fraction ($\epsilon_e$) of the post-shock
energy density goes to electrons, one has
$\gamma_m=(m_p/m_e)\epsilon_e[(p-2)/(p-1)]\gamma=610\epsilon_eg_p\gamma$, where
$g_p=3(p-2)/(p-1)$.

If the cooling time of electrons with Lorentz factor of $\gamma_m$ is shorter
than $t$ (i.e. $\gamma_m>\gamma_c$), the steady electron energy distribution
turns out to be $dN_e/d\gamma_e\propto \gamma_e^{-2}$ for
$\gamma_c<\gamma_e<\gamma_m$ and $dN_e/d\gamma_e\propto \gamma_e^{-(p+1)}$ for
$\gamma_e\ge \gamma_m$. In such a fast cooling regime, the spectral luminosity
at $\epsilon_\gamma$ is (e.g. Blumenthal \& Gould 1970)
\begin{equation}
L_{\epsilon_\gamma}=\left \{ \begin{array}{ll} L_{\epsilon_\gamma,{\rm
max}}(\epsilon_\gamma/\epsilon_c)^{-1/2}, &
{\epsilon_c<\epsilon_\gamma<\epsilon_m},\\ L_{\epsilon_\gamma,{\rm
max}}(\epsilon_m/\epsilon_c)^{-1/2}
(\epsilon_\gamma/\epsilon_m)^{-p/2}, & {\epsilon_\gamma\ge\epsilon_m},
\end{array} \right.
\end{equation}
where $\epsilon_c\simeq \gamma_c^2\epsilon_0$ and $\epsilon_m\simeq
\gamma_m^2\epsilon_0$.

On the other hand, if $\gamma_m<\gamma_c$ (slow cooling), the steady electron
energy distribution becomes $dN_e/d\gamma_e\propto \gamma_e^{-p}$ for
$\gamma_m<\gamma_e<\gamma_c$ and $dN_e/d\gamma_e\propto \gamma_e^{-(p+1)}$ for
$\gamma_e\ge \gamma_c$. In this case, the spectral luminosity at
$\epsilon_\gamma$ reads
\begin{equation}
L_{\epsilon_\gamma}=\left \{ \begin{array}{ll} L_{\epsilon_\gamma,{\rm
max}}(\epsilon_\gamma/\epsilon_c)^{-(p-1)/2}, &
{\epsilon_m<\epsilon_\gamma<\epsilon_c},\\ L_{\epsilon_\gamma,{\rm
max}}(\epsilon_c/\epsilon_m)^{-(p-1)/2}
(\epsilon_\gamma/\epsilon_c)^{-p/2}, & {\epsilon_\gamma\ge\epsilon_c}.
\end{array} \right.
\end{equation}

\section{Shock dynamics and emission luminosity}

We assume that the luminosity of a relativistic outflow is $L_w(t)\propto t^k$
for $t<t_w$ and $L_w(t)\propto t^{-5/3}$ for $t\ge t_w$, as is suggested by the
popular collapsar model for long-duration, soft-spectrum GRBs (MacFadyen,
Woosley \& Heger 2001). The Lorentz factor of the outflow is $\gamma_w\gg 1$.
The comoving proton number density in the outflow rest frame is $n_w=L_w/(4\pi
R^2\gamma_w^2m_pc^3)$. For simplification, we also assume that a preexisting
shell is so slow (as compared with the outflow) that the shell can be
considered as being at rest, and its uniform proton number density is $n$. We
assume that the collision happens at an initial radius $R_i$. Similar to Sari
\& Piran (1995), the outflow-shell interaction beyond $R_i$ could be described
through two shocks, a reverse shock that propagates into the outflow, and a
forward shock that propagates into the shell. There are four regions separated
by the two shocks, (1) the unshocked shell, (2) the forward-shocked shell, (3)
the reverse-shocked outflow, and (4) the unshocked outflow. The shock dynamics
depends on whether the reverse shock has crossed the outflow. It is clear that
at two stages, i.e., during and after the reverse-shock crossing, the shock
dynamics is different and thus the emission luminosity should also be
different. In the following, we discuss these two stages separately.

\subsection{During the reverse-shock crossing}

As shown by Sari \& Piran (1995), if $\gamma_w^2\gg n_w/n$, the reverse shock
is relativistic and the Lorentz factor of regions 2 \& 3 measured in the rest
frame of region 1 is
\begin{equation}
\gamma_{\rm fs}=\left({\gamma_w\over2}\right)^{1/2}\left({n_w\over
n}\right)^{1/4}=3.24L_{w,50}^{1/8}n_9^{-1/8}(t_3+t_{i,3})^{-1/4},
\end{equation}
where the convention $Q=10^x\times Q_x$ in cgs units has been adopted and
$t_i=t_{i,3}\times 10^3\,{\rm s}\equiv R_i/(2\gamma_{\rm fs}^2c)$. The shock
radius becomes $R=2\gamma_{\rm fs}^2ct+R_i\equiv 2\gamma_{\rm
fs}^2c(t+t_i)=6.3\times
10^{14}L_{w,50}^{1/4}n_9^{-1/4}(t_3+t_{i,3})^{1/2}$\,cm. In addition, the
Lorentz factor of region 3 measured in the rest frame of region 4 is
\begin{equation}
\gamma_{\rm rs}=\left({\gamma_w\over2}\right)^{1/2}\left({n_w\over
n}\right)^{-1/4}=1.54L_{w,50}^{-1/8}n_9^{1/8}\gamma_{w,1}(t_3+t_{i,3})^{1/4}.
\end{equation}

From eq. (5), we calculate the cooling Lorentz factor of the relativistic
electrons
\begin{equation}
\gamma_c=2.0L_{0,46}^{-1}L_{w,50}^{5/8}n_9^{-5/8}t_3^{-1}(t_3+t_{i,3})^{3/4}.
\end{equation}
The characteristic Lorentz factors of the relativistic electrons in regions 2
\& 3 are, respectively,
\begin{equation}
\gamma_m=\left \{
       \begin{array}{ll}
       20\epsilon_{e,-2}g_pL_{w,50}^{1/8}n_9^{-1/8}(t_3+t_{i,3})^{-1/4},
       & {{\rm region}\,\,2},\\
       10\epsilon_{e,-2}g_pL_{w,50}^{-1/8}n_9^{1/8}\gamma_{w,1}(t_3+t_{i,3})^{1/4},
       & {{\rm region}\,\,3}.
       \end{array} \right.
\end{equation}
for a soft photon source with a black body spectrum and a typical photon energy
$\epsilon_0\sim k_BT\sim 0.16$\,keV (as in GRB 060218), the cooling energy of
the scattered photons is
\begin{equation}
\epsilon_c=
0.64L_{0,46}^{-2}L_{w,50}^{5/4}n_9^{-5/4}\zeta_Tt_3^{-2}(t_3+t_{i,3})^{3/2}\,{\rm
keV},
\end{equation}
where $\zeta_T=k_BT/0.16\,{\rm keV}$. The characteristic energies of the
scattered photons from regions 2 \& 3 are, respectively,
\begin{equation}
\epsilon_m=\left \{
       \begin{array}{ll}
       64\epsilon_{e,-2}^2g_p^2L_{w,50}^{1/4}n_9^{-1/4}\zeta_T
       (t_3+t_{i,3})^{-1/2}\,{\rm keV}, & {{\rm region}\,\,2},\\
       16\epsilon_{e,-2}^2g_p^2L_{w,50}^{-1/4}n_9^{1/4}\gamma_{w,1}^2
       \zeta_T(t_3+t_{i,3})^{1/2}\,{\rm keV}, & {{\rm region}\,\,3}.
       \end{array} \right.
\end{equation}
The total electron numbers in regions 2 \& 3 are $N_{e,2}=4\pi R^2(2\gamma_{\rm
fs}^2ct)n= 3.1\times 10^{54}L_{w,50}^{3/4}n_9^{1/4}t_3(t_3+t_{i,3})^{1/2}$ and
$N_{e,3}=N_{e,2}(\gamma_{\rm rs}/\gamma_{\rm fs})=1.5\times
10^{54}L_{w,50}^{1/2}n_9^{1/2}\gamma_{w,1}t_3(t_3+t_{i,3})$, respectively. From
eq. (4), we then obtain the peak spectral luminosities for regions 2 \& 3,
\begin{equation}
L_{\epsilon_\gamma,{\rm max}}=\left \{ \begin{array}{ll} 2.7\times
10^{46}L_{0,46}L_{w,50}^{1/4}n_9^{3/4}\zeta_T^{-1}t_3(t_3+t_{i,3})^{-1/2}
\,\,{\rm erg\,keV^{-1}\,s^{-1}}, & {{\rm region}\,\,2},\\ 1.3\times
10^{46}L_{0,46}n_9\gamma_{w,1}\zeta_T^{-1}t_3 \,\,{\rm
erg\,keV^{-1}\,s^{-1}}, & {{\rm region}\,\,3}.  \end{array} \right.
\end{equation}
According to equations (12)-(14), the spectral luminosities at scattered photon
energy $\epsilon_\gamma$ in regions 2 \& 3 are
\begin{equation}
L_{\epsilon_\gamma}=\left \{ \begin{array}{ll} 2.2\times
10^{46}L_{w,50}^{7/8}n_9^{1/8}\zeta_T^{-1/2}(t_3+t_{i,3})^{1/4}
\epsilon_{\gamma,1{\rm keV}}^{-1/2} \,\,{\rm erg\,keV^{-1}\,s^{-1}}, &
{{\rm region}\,\,2},\\ 1.0\times
10^{46}L_{w,50}^{5/8}n_9^{3/8}\gamma_{w,1}\zeta_T^{-1/2}(t_3+t_{i,3})^{3/4}
\epsilon_{\gamma,1{\rm keV}}^{-1/2} \,\,{\rm erg\,keV^{-1}\,s^{-1}}, &
{{\rm region}\,\,3}, \end{array} \right.
\end{equation}
for $\epsilon_c<\epsilon_\gamma<\epsilon_m$, and
\begin{equation}
L_{\epsilon_\gamma}=\left \{ \begin{array}{ll} 4.9\times
10^{47}\epsilon_{e,-2}^{p-1}g_p^{p-1}L_{w,50}^{(p+6)/8}n_9^{-(p-2)/8}
\zeta_T^{(p-2)/2}(t_3+t_{i,3})^{-(p-2)/4}\epsilon_{\gamma,1{\rm
keV}}^{-p/2} \,\,{\rm erg\,keV^{-1}\,s^{-1}}, & {{\rm region}\,\,2},\\
7.5\times
10^{46}\epsilon_{e,-2}^{p-1}g_p^{p-1}L_{w,50}^{(6-p)/8}n_9^{(p+2)/8}\gamma_{w,1}^p
\zeta_T^{(p-2)/2}(t_3+t_{i,3})^{(p+2)/4} \epsilon_{\gamma,1{\rm
keV}}^{-p/2} \,\,{\rm erg\,keV^{-1}\,s^{-1}}, & {{\rm region}\,\,3},
\end{array} \right.
\end{equation}
for $\epsilon_c<\epsilon_m<\epsilon_\gamma$. Here $\epsilon_{\gamma,1{\rm
keV}}=\epsilon_\gamma/1\,{\rm keV}$, and the coefficients in eq. (16) have been
calculated for $p=2.5$.

\subsection{After the reverse-shock crossing}

We assume that the reverse-shock crossing time is $t_w\sim t_i$. After this
time, the central engine luminosity is insignificant because it decays as
$L_w\propto t^{-5/3}$ (MacFadyen et al. 2001). Although the electrons in region
3 have cooled down, the higher-latitude scattered photons would reach the
observer at a later time.  This is the well-known curvature effect (Kumar \&
Panaitescu 2000; Zhang et al.  2006; Wu et al. 2006; Liang et al. 2006). In the
relativistic limit, this effect leads to the emission luminosity
$L_{\epsilon_\gamma}\propto t^{-(2+\beta)}$ (where $\beta$ is the spectral
index). As argued by Zhang et al. (2006), the temporal index $\alpha=2+\beta$
conclusion is essentially unchanged by deceleration of region 3. Meanwhile, the
forward shock still sweeps up its surrounding shell and its Lorentz factor
decreases as $\gamma_{\rm fs}\propto t^{-3/8}$. The cooling Lorentz factor
decays as $\gamma_c\propto(\gamma_{\rm fs}U'_0t)^{-1}\propto\gamma_{\rm
fs}R^2t^{-1}\propto t^{-7/8}$ (see eq.[5]) and the cooling energy of the
scattered photons deccay as $\epsilon_c\propto t^{-7/4}$. The peak spectral
luminosity increases as $L_{\epsilon_\gamma,{\rm max}}\propto N_eR^{-2}\propto
R\propto t^{1/4}$. Thus the spectral luminosity at
$\epsilon_c<\epsilon_\gamma<\epsilon_m$ decays as
$L_{\epsilon_\gamma}=L_{\epsilon_\gamma,{\rm
max}}(\epsilon_\gamma/\epsilon_c)^{-1/2}\propto t^{-5/8}$.

\section{Application to GRB 060218}

The prompt emission light curves detected by BAT and XRT are presented in
Campana et al. (2006): The 15-150 keV luminosity and the 0.3-10 keV luminosity
$\sim 900$ s after the burst increases as $L_{\rm BAT}\propto
t^{\alpha_{\gamma,1}}$ with $\alpha_{\gamma,1}\sim 0.3$ and $L_{\rm XRT}\propto
t^{\alpha_{x,1}}$ with $\alpha_{x,1}\sim 0.7$, respectively. Subsequently the
15-150 keV luminosity decays as $L_{\rm BAT}\propto t^{\alpha_{\gamma,2}}$ with
$\alpha_{\gamma,2}\sim -4.5$, while the 0.3-10 keV luminosity declines slowly
as $L_{\rm XRT}\propto t^{\alpha_{x,2}}$ with $\alpha_{x,2}$ with
$\alpha_{x,2}\sim -0.6$ until $t\sim 2600$ s, after which the lightcurve
declines rapidly as $L_{\rm XRT}\propto t^{\alpha_{x,3}}$ with
$\alpha_{x,3}\sim -4.0$. The different light curves detected by BAT and XRT
suggest that their corresponding emission could have originated from different
regions. In the following, we show that the X-ray emission and $\gamma$-ray
emission would be produced from regions 2 and 3 in our model, respectively. An
example fit to the observational data with our model is displayed in Figure 1.

We first consider the emission luminosity at the peak time $t_{\rm peak}\sim
900$ s. Taking $p\sim 2.5$, $t_i\sim 10^3$ s and $k_BT\sim 0.16$ keV, we
calculate the $15-150$ keV luminosity by using the second expression of eq.
(16),
\begin{equation}
L_{\rm BAT}=\int_{15\,{\rm keV}}^{150\,{\rm
keV}}L_{\epsilon_\gamma}d\epsilon_\gamma\sim 3.3\times
10^{47}\epsilon_{e,-2}^{1.5}L_{w,50}^{7/16}n_9^{9/16}\gamma_{w,1}^{2.5}\,{\rm
erg\,s^{-1}}\sim 1.8\times 10^{46}\,{\rm erg\,s^{-1}}.
\end{equation}
Using the first expression of eq. (15), we have the $0.3-10$ keV
luminosity
\begin{equation}
L_{\rm XRT}=\int_{0.3\,{\rm keV}}^{10\,{\rm
keV}}L_{\epsilon_\gamma}d\epsilon_\gamma\sim 1.7\times
10^{47}L_{w,50}^{7/8}n_9^{1/8}\,{\rm erg\,s^{-1}}\sim 3.3\times
10^{46}\,{\rm erg\,s^{-1}}.
\end{equation}
From eqs. (17) and (18), we find the shell number density and the outflow
luminosity at the peak time $t_{\rm peak}\simeq t_w$,
\begin{equation}
n_9\sim 0.02\epsilon_{e,-2}^{-3}\gamma_{w,1}^{-5},
\end{equation}
\begin{equation}
L_{w,50}\sim 0.3\epsilon_{e,-2}^{3/7}\gamma_{w,1}^{5/7},
\end{equation}
respectively.

Second, we discuss the light curves before $t_{\rm peak}$. Theoretically, the
$\gamma$-ray and X-ray luminosities increase as $L_{\rm BAT}\propto
L_w^{(6-p)/8}$ and $L_{\rm XRT}\propto L_w^{7/8}$, respectively. Considering
$L_w\propto t^k$ before $t_w$, we require $\alpha_{\gamma,1}\simeq (6-p)k/8\sim
0.3$ and $\alpha_{x,1}\simeq 7k/8\sim 0.7$, giving $k\sim 0.8$. This indicates
that the outflow luminosity increases slowly with time before $t_{\rm peak}$.
In the collapsar model, numerical simulations by MacFadyen et al. (2001)
indicate that the accretion rate of the fallback disk by the central
stellar-mass black hole typically shows a slowly-rising behavior before the
well-known $\dot{M}\propto t^{-5/3}$ accretion starts. The outflow luminosity
has a similar temporal power because of a nearly constant Blandford-Znajek
power efficiency for the rotating black hole (McKinney 2005). Thus, our fit is
consistent with the prediction of the collapsar model.

Third, after $t_w$, the reverse shock has crossed the outflow and the curvature
effect of region 3 leads to the sharp $\gamma$-ray luminosity decay, i.e.
$L_{\rm BAT}\propto t^{-(2+\beta)}\sim t^{-3.25}$. This is marginally
consistent with the steep decay of the $\gamma$-ray emission. A larger value of
$p$ or a potential spectral break at higher energies would make the curvature
effect more consistent with the observed decay. Meanwhile, region 2 could have
entered the self-similar solution of Blandford \& McKee (1976) and its X-ray
luminosity decays as $L_{\rm XRT}\propto t^{-5/8}$. This is in good agreement
with the observed value of $\alpha_{x,2}$.

Fourth, we note from Campana et al. (2006) that after $\sim 2600$ s both the
non-thermal component luminosity and the thermal component luminosity decayed
rapidly. This steep decay is the result of the combination of the rapid decay
of the cocoon emission (Pe'er et al. 2006) and the curvature effect of region 2
after the forward shock has crossed the shell.

Finally, the forward shock, after crossing the slow shell, would sweep into the
circumburst stellar wind and produce a normal afterglow as observed by {\em
Swift} XRT. A single power law light curve of this afterglow indicates that the
outflow is not strongly collimated (Fan, Piran \& Xu 2006).

\section{Conclusions}

The abundant multiwavelength prompt emission data of GRB 060218/SN 2006aj make
it possible to constrain the prompt emission mechanisms and model parameters in
great detail. The data clearly disfavor the conventional internal
shock/synchrotron radiation models that predict strong prompt UV optical
emission. Instead the data are consistent with a model that invokes inverse
Compton scattering off the thermal photons due to shock breakout. The chromatic
lightcurve features of the prompt emission require that the prompt BAT-band
emission and XRT-band emission are produced in two different regions, and we
identify these as a pair of shocks upon the interaction between a relativistic
outflow and a preexisting slow shell. Fitting the data with model poses
constraints on the physical parameters of the outflow and the shell, in
particular, requires a central engine wind with a slowly-increasing luminosity.
This behavior is consistent with the prediction of the popular collapsar model.

\acknowledgments This work was supported by NASA under grants NNG05GB67G,
NNG05GH92G and NNG05GH91G (BZ, ZGD \& EWL) and the National Natural
Science Foundation of China under grants 10221001 \& 10233010 (ZGD)
and 10463001 (EWL).

\clearpage
\begin{figure}
\plotone{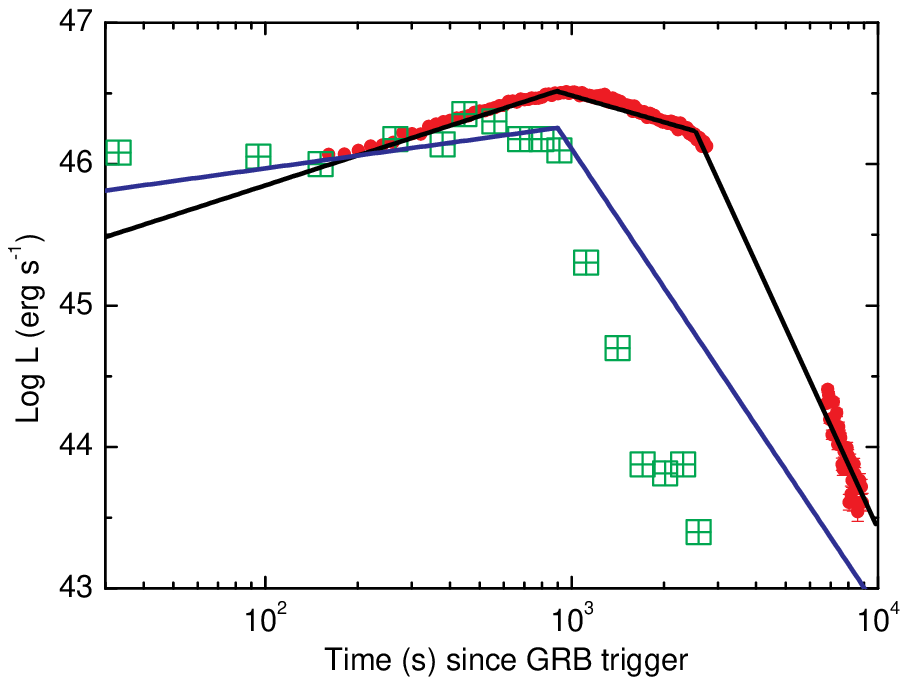} \caption{An example fit to the prompt emission data of GRB
060218 with our model, for $p\sim 2.5$, $k\sim 0.8$, and $t_w\sim 900$ s. The
red dots and green squares represent the XRT and BAT data (taken from version 2
of Campana et al. 2006), respectively. In our fit, the required peak outflow
luminosity is $L_{w,50}\sim 0.3\epsilon_{e,-2}^{3/7}\gamma_{w,1}^{5/7}$, the
shell number density is $n_9\sim 0.02\epsilon_{e,-2}^{-3}\gamma_{w,1}^{-5}$,
and the shell width is $\Delta R_{\rm shell}\sim 3.2\times
10^{15}\epsilon_{e,-2}^{6/7}\gamma_{w,1}^{10/7}$ cm. The rising segments of our
theoretical $\gamma$-ray and X-ray light curves are $L_{\rm BAT}\propto
t^{(6-p)k/8}$ and $L_{\rm XRT} \propto t^{7k/8}$, respectively. During the
period between $\sim 900$ to $2600$ s, the decaying index of the X-ray
luminosity is $-5/8$ and the decaying index of the $\gamma$-ray luminosity
becomes $-(2+\beta)$ because of the curvature effect after the reverse shock
has crossed the outflow. At $\sim 2600$ s, the forward shock has crossed the
shell, and hence, the X-ray luminosity shows a rapid decay due to the curvature
effect. However, the actual decay becomes steeper because of the rapid decrease
of the observed thermal luminosity. \label{fig1}}
\end{figure}

\end{document}